\documentclass[aps,prl,twocolumn,superscriptaddress]{revtex4-1}

\usepackage{graphicx}

\begin{document}

\title{Electron Heating by Debye-Scale Turbulence in Guide-Field Reconnection}


\author{Yu. V. Khotyaintsev}
\email[]{yuri@irfu.se}
\affiliation{Swedish Institute of Space Physics, Uppsala, Sweden}

\author{D. B. Graham}
\affiliation{Swedish Institute of Space Physics, Uppsala, Sweden}

\author{K. Steinvall}
\affiliation{Swedish Institute of Space Physics, Uppsala, Sweden}

\author{L. Alm}
\affiliation{Swedish Institute of Space Physics, Uppsala, Sweden}

\author{A. Vaivads}
\affiliation{Department of Space and Plasma Physics, KTH Royal Institute of Technology, Stockholm, Sweden}

\author{A. Johlander}
\affiliation{Univ. of Helsinki, Helsinki, Finland}

\author{C. Norgren}
\affiliation{Univ. of Bergen, Bergen, Norway}

\author{W. Li}
\affiliation{State Key Laboratory of Space Weather, National Space Science Center, Chinese Academy of Sciences, Beijing, China}

\author{A. Divin}
\affiliation{Physics Department, St. Petersburg State University, St. Petersburg, Russia}


\author{H. S. Fu}
\affiliation{School of Space and Environment, Beihang University, Beijing, China}

\author{K.-J. Hwang}
\affiliation{Southwest Research Institute, San Antonio, TX, USA}


\author{N. Ahmadi}
\affiliation{Laboratory of Atmospheric and Space Physics, University of Colorado, Boulder, CO, USA}

\author{O. Le Contel}
\affiliation{Laboratoire de Physique des Plasmas, CNRS/Ecole Polytechnique/Sorbonne Universit\'e/Universit\'e Paris-Sud/Observatoire de Paris, Paris, France}

\author{D. J. Gershman}
\affiliation{NASA Goddard Space Flight Center, Greenbelt, MD, USA}

\author{C. T. Russell}
\affiliation{University of California, Los Angeles, CA, USA}

\author{R. B. Torbert}
\affiliation{University of New Hampshire, Durham, NH, USA}


\author{J. L. Burch}
\affiliation{Southwest Research Institute, San Antonio, TX, USA}


\date{\today}

\begin{abstract}
We report electrostatic Debye-scale turbulence developing within the diffusion region of asymmetric magnetopause reconnection with moderate guide field using observations by the Magnetospheric Multiscale (MMS) mission. We show that Buneman waves and beam modes cause efficient and fast thermalization of the reconnection electron jet by irreversible phase mixing, during which the jet kinetic energy is transferred into thermal energy. Our results show that the reconnection diffusion region in the presence of a moderate guide field is highly turbulent, and that electrostatic turbulence plays an important role in electron heating.
\end{abstract}


\maketitle


The role of turbulence in reconnection is a subject of active debate. Waves are frequently observed in association with reconnection and have been suggested to play important roles in reconnection \citep{fujimoto2011, fu2017, khotyaintsev2019}.
For antiparallel and weak-guide-field cases, the wave activity is mostly found in the separatrix region, which is a kinetic boundary separating the inflow and outflow regions \citep{lindstedt2009}. Electrons are accelerated in this region by an electrostatic potential \citep{fujimoto2014, egedal2015}. The resulting fast electron streaming can generate a variety of plasma waves, including electron holes, Langmuir waves, Buneman and beam modes, and whistlers \citep{lapenta2011, divin2012a, viberg2013}. Wave-particle interactions can lead to thermalization of streaming electrons \citep{fujimoto2014,holmes2019}.
 While in the electron diffusion region (EDR) the electron dynamics is largely laminar, and dominated by electron meandering motion \citep{chen2016,torbert2018}.

For larger guide fields the magnetic field at the X-line does not vanish, and thus the effects of meandering are reduced. The electron current in the EDR flows along the guide field. Numerical simulations suggest that in such situations streaming instabilities lead to development of kinetic turbulence over a broad frequency range in the EDR vicinity \citep{munoz2018}, with Buneman and two-streaming instabilities being responsible for the high-frequency (above the lower-hybrid frequency) fluctuations. These instabilities can lead to electron heating, anomalous resistivity and potentially increase of the reconnection rate \citep{drake2003, che2011, che2013, che2017}. Buneman waves in the EDR vicinity have  been reported using recent Magnetospheric Multiscale (MMS) observations \citep{khotyaintsev2016}, but the overall role of the turbulence and streaming instabilities for the EDR physics requires observational verification. In this letter, we use MMS to investigate electrostatic (ES) turbulence in the reconnection diffusion region and its effect on electron dynamics. We show that large-amplitude ES turbulence is observed at the X-line and it strongly affects the electron jet, leading to fast thermalization of electrons.

\begin{figure}
\includegraphics[width=8.6cm]{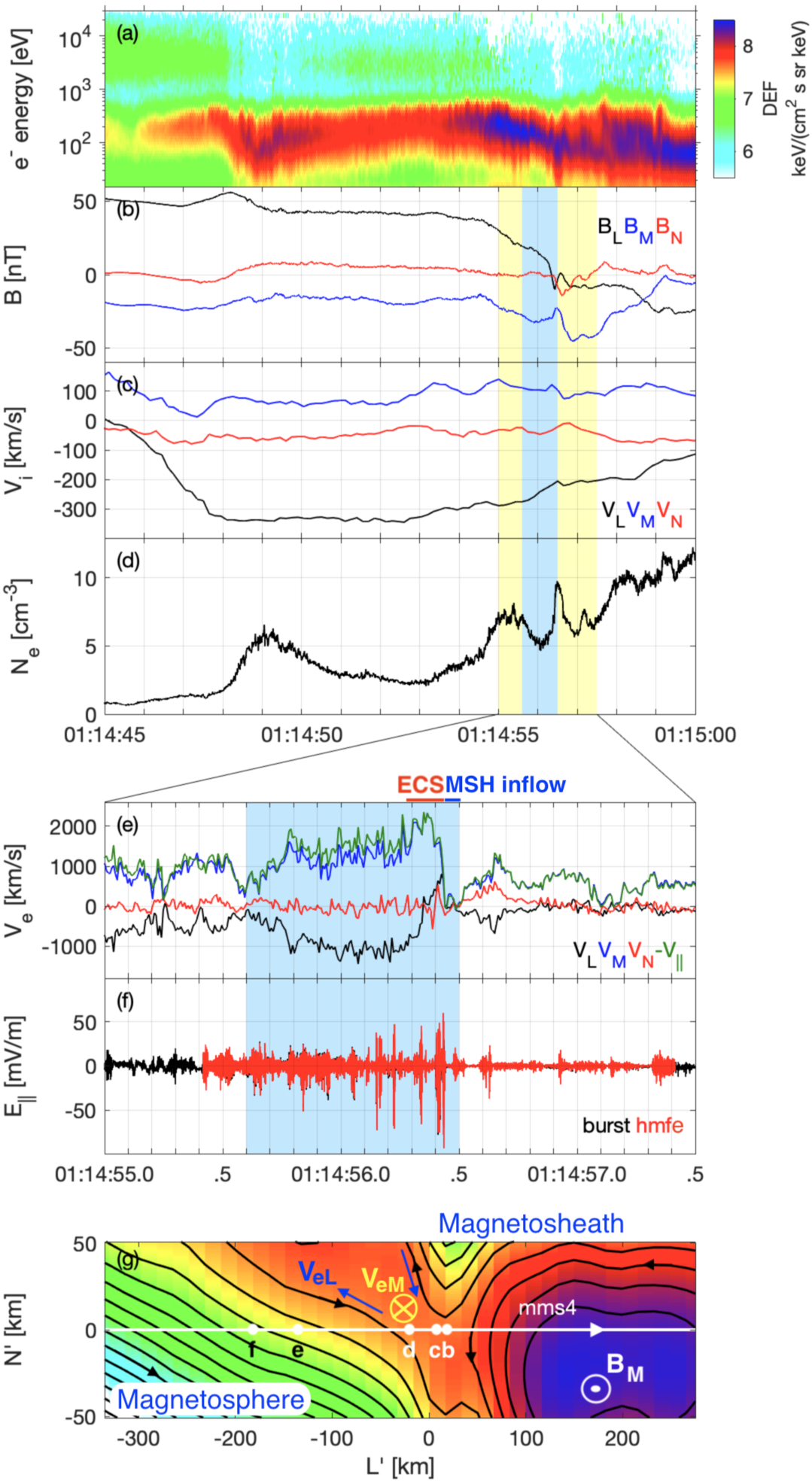}
\caption{Top: Overview of magnetopause crossing on December 2, 2015. (a) Electron spectrogram measured by FPI \citep{pollock2016}, (b) magnetic field from FGM \citep{russell2014}, (c) ion velocity and (d) electron density from FPI. Middle: Electron current sheet (ECS). (e) Electron velocity from FPI and (f) $E_{||}$ measured by EDP\citep{lindqvist2014,ergun2014}, snapshot of the highest resolution HMFE data (65 kHz sampling, red) plotted on top of the burst data (16 kHz sampling, black). Electron moments are sampled at 7.5 ms cadence \citep{rager2018}. Bottom: Grad-Shafranov reconstruction showing crossing through the X-line. Points f-b along the spacecraft trajectory mark the distributions in Fig.\ref{fig:dists}. The color indicates the amplitude of the vector potential.  Reconstruction has been performed over the time interval of panels (e) and (f). Blue/yellow arrows indicate directions of the electron flow.
\label{fig:over}}
\end{figure}

We analyze an EDR crossing by MMS on December 2, 2015, Fig.~\ref{fig:over}. 
The four MMS spacecraft were separated on average by 10\,km $\sim6.5 d_e$, where the electron inertial length $d_e=c/\omega_{pe}=1.5$ km (using the magnetosheath density of 12 cm$^{-3}$). MMS 3 and 4 were separated by $\sim1 d_e$ in the direction normal to the magnetopause (MP) and both detected similar EDR signatures; below we show data from MMS4. Boundary LMN coordinates were obtained using minimum variance analysis (MVA) of the magnetic field $\mathbf{B}$ and current $\mathbf{J}$,  L =  [0.02 -0.57 0.82] and N = [0.86    -0.41    -0.31] in geocentric solar ecliptic (GSE) coordinates, $\mathbf{M}=\mathbf{N}\times \mathbf{L}$. 
MMS crossed the MP boundary from the low-density high-temperature magnetospheric to high-density low-temperature magnetosheath sides (seen as the sharp change in electron energy spectrum in Fig.~\ref{fig:over}a and density in Fig.~\ref{fig:over}d). The MP current layer can be seen as the reversal of $B_L$, Fig.~\ref{fig:over}b. 

A background $B_M \sim$ 20\,nT is observed throughout the interval, corresponding to a guide field of 50\% (100\%) of the magnetospheric (magnetosheath) $B_L$. We observe an ion jet which is tangential to the boundary ($V_L$ dominant) on the low-density side of the MP, Fig.~\ref{fig:over}c. The onset of the strong ion flow at 01:14:48\,UT coincides with plasma density increase (Fig.~\ref{fig:over}d), as well as with a decrease in flux of energetic ($>$1\,keV) electrons; we interpret this boundary as the magnetospheric separatrix. Between the separatrix and the MP crossing we observe the separatrix region, characterized by the Hall electric field and electrostatic lower-hybrid drift instability (LHDI) turbulence \citep{khotyaintsev2016,graham2019}.

We focus on the region around the neutral point, $B_L \sim 0$, where fast electron flows are observed (Fig.~\ref{fig:over}e). The electron bulk speed peak, $v_e\sim$ 2300\,km/s, corresponds to electron Mach number $M_{e\perp} = v_e/v_{Te\perp} = 0.55$.
At $B_L=0$ the electron jet is in the $M$ direction, i.e. directed out of the reconnection plane. This jet is predominantly aligned with $\mathbf{B}$. The peak out-of-plane electron jet, $|v_{e,M}|\simeq$2000 km/s, lasts for 0.1 sec, which corresponds to a spatial scale in the N direction of 7.5 km $\sim5 d_e$, thus we are observing an electron-scale current sheet (ECS). Here we have used the boundary normal velocity $v_N = $ -75 km/s, determined from multi-spacecraft timing of $n_e$, $T_e$, and $E_n$ which is consistent with the observed $v_{iN}$. 
Within the ECS we observe non-gyrotropic crescent distributions \citep{burch2015,norgren2016} (not shown). We also observe  large-amplitude fluctuating $E_{||}$, Fig.~\ref{fig:over}f, which indicates possible instability of the fast electron jet.

Figure \ref{fig:over}g shows a two-dimensional (2D) Grad-Shafranov reconstruction of the magnetic topology near the X-line. The reconstruction was performed in the L-N plane, assuming the structure is invariant along the M-direction. The reconstruction is performed in the co-moving frame of the magnetic structure, where it can be assumed that it is approximately time stationary \citep{hau1999}. This velocity was determined through multi-spacecraft timing analysis on the magnetic field. This frame is then rotated so that the path of the spacecraft follows the X-axis of the reconstruction box, at X=0. With a $v_N=-75$ km/s and $v_L=-250$ km/s, the L' and N' axes of the Grad-Shafranov reconstruction differ from the L and N directions by approximately 17 degrees. The reconstruction indicates X-line magnetic topology in the vicinity of the ECS. However, there is no magnetic null at the X-line, because of the finite guide field, $B_M$. This topology is confirmed by the FOTE analysis \citep{fu2015} (not shown). The observation of the ECS with a high Mach number electron flow, $M_{e\perp} \lesssim 1$ at the X-line, as well as of crescent distributions indicate that MMS4 is located in the EDR vicinity. 

The change of $v_{eL}$ sign at the ECS (Fig.~\ref{fig:over}e) is consistent with the ECS crossing in the N direction (switch of the L-flow away from the X-line to towards the X-line illustrated in Fig.~\ref{fig:over}g). Following the positive $v_{eL}$ interval, the electron flow reduces to zero, which we interpret as transition to the inflow region on the high density side. This is confirmed by a brief dropout in energetic (magnetospheric) electrons (Fig.~\ref{fig:over}a), indicating no magnetic field connection to the magnetospheric side. After this the spacecraft encounter an ion-scale flux rope ($L'>50$ km in Fig.~\ref{fig:over}g). 

\begin{figure}
\includegraphics[width=8.6cm]{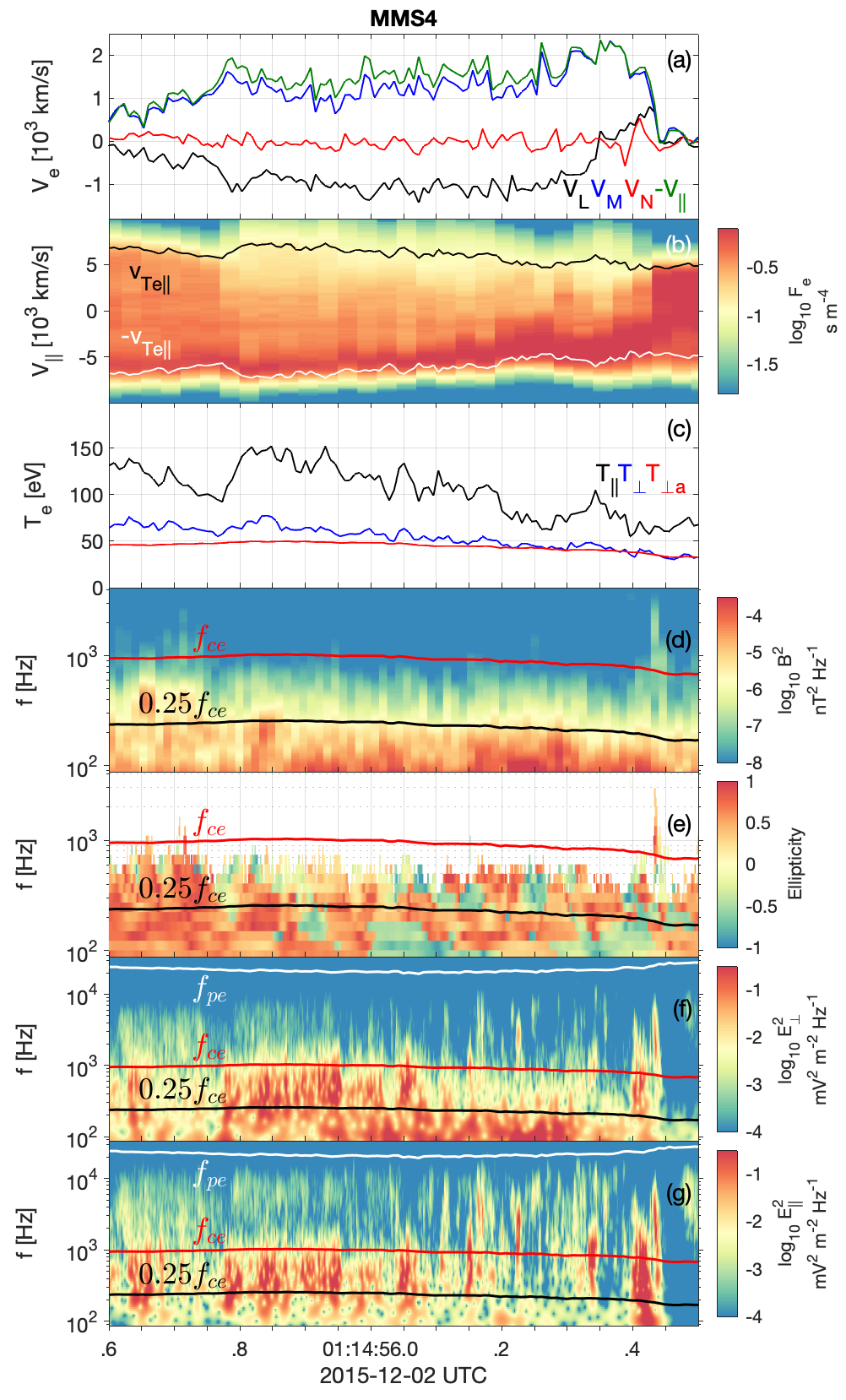}
\caption{Electron jet and associated waves. (a) Electron velocity, (b) integrated 1D velocity distribution function (VDF) $f(v_{||})$ based on 30 ms FPI distributions showing electron beam (jet), 
(c) electron temperatures $T_{e||}$, $T_{e\perp}$, and temperature expected from adiabatic betatron heating $T_{\perp a}$ (d,e) B spectrum and ellipticity, (f, g)  $E_{\perp}$ and $E_{||}$ spectrum. $f_{ce}$ and $f_{pe}$ stand for electron-cyclotron and electron-plasma frequencies, respectively.
\label{fig:waves}}
\end{figure}

As one can see from Fig.~\ref{fig:over}g, prior to the X-line encounter MMS4 is moving primarily tangentially to the boundary, spending significant time within the jet region (shaded area in Fig.~\ref{fig:over}e and \ref{fig:over}f). 
The evolution of the reduced one-dimensional (1D) electron velocity distribution functions (VDFs), $f_e(v_{||})$, in this region are shown in Fig.~\ref{fig:waves}b. Such reduced VDFs are convenient as they capture the relevant electron dynamics, which is predominantly field aligned in the guide-field case. In the beginning of the interval, before 01:14:55.8 UT, $f_e(v_{||})$ is symmetric, indicating the electrons are largely trapped in the field-aligned direction. After this, $f_e(v_{||})$ becomes asymmetric, with a narrow anti-field aligned beam (originating from the high-density side of the boundary) on top of a more energetic counter-streaming population (of magnetospheric/low-density side origin). Closer to the X-line, at 01:14:56.2 UT, the beam becomes slower and more spread in energy corresponding to a plateau in $f_e(v_{||})$. Finally, in the inflow region, $v_e \sim 0$, the distribution is again symmetric. So, within the jet we observe VDFs characteristic for the reconnected field-lines: dense and cold magnetosheath population is mixing with hot magnetospheric population. We will show that this is not a simple mixing, but it is affected by parallel electric fields leading to electron acceleration as well as by waves trapping and scattering the electrons.

\begin{figure}
\includegraphics[width=8.6cm]{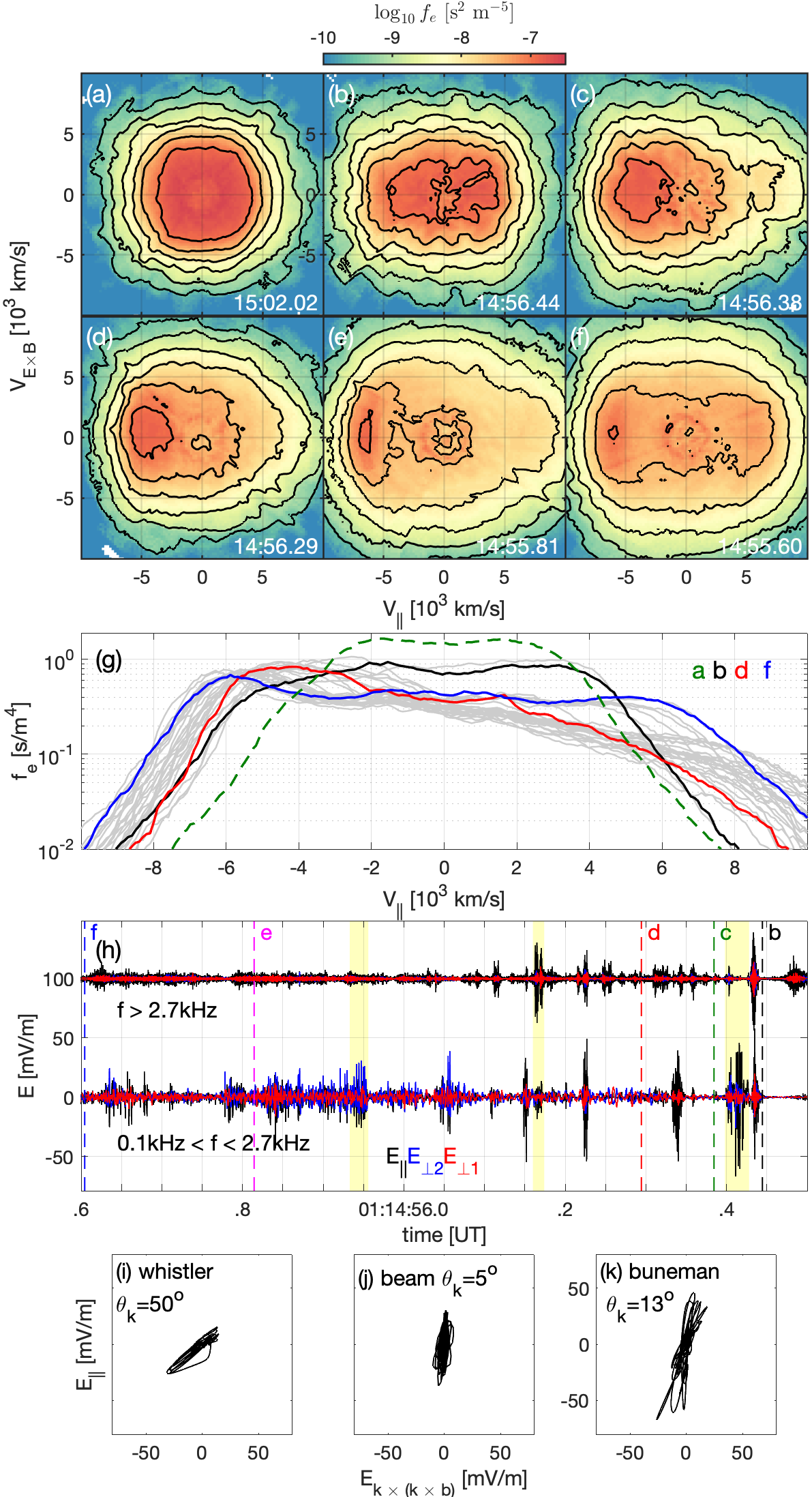}
\caption{Detailed electron distributions and associated $E_{||}$ waves. (a)-(f) 2D reduced electron VDFs observed at the times indicated in the panels (based on 30 ms FPI distributions), (g) integrated 1D VDFs, the colored lines correspond to the 2D distributions above, and the grey lines show all the other distributions during this time interval, (h) $E$ - HF and LF waveforms, (i)-(k) typical wave E polarization for (i) oblique whistler, sampled at 01:14:55.95 UT, (j) beam mode, sampled at 01:14:56.17 UT, and (k) Buneman mode, sampled at 01:14:56.41 UT.
\label{fig:dists}}
\end{figure}

Selected reduced 2D VDFs $f_e(v_{||}, v_{\perp})$ within the jet from the magnetosheath to the magnetospheric sides are shown in Fig.~\ref{fig:dists}b-f, and the corresponding 1D distributions are provided in Fig.~\ref{fig:dists}g. For reference, we also include the magnetosheath distribution further away from the reconnection site (Fig.~\ref{fig:dists}a). The distribution at the high density separatrix, Fig.~\ref{fig:dists}b, is stretched in the parallel direction, but squeezed in the perpendicular direction. Immediately after the separatrix crossing (Fig.~\ref{fig:dists}c) only the $v_{||}<0$ part of the distribution remains (moving towards the X-line), and at the same time a narrow field-aligned energetic population at positive $v_{||}>5000$ km/s appears, which is of magnetospheric origin. This indicates the magnetic field lines connect to the magnetosheath on one side and magnetosphere on the other. Applying Liouville mapping of distribution (a) to (c) in a similar way to Ref. \citep{eriksson2018}, i.e. assuming the source electrons (a) are accelerated along a magnetic flux tube by $E_{||}$, we find the net accelerating potential $\Delta \Phi_{||} = - \int E_{||} \mathrm{d}l \sim80$ V $\sim T_e$. The distribution in Fig.~\ref{fig:dists}d is then further accelerated with respect to Fig.~\ref{fig:dists}c, and in Fig.~\ref{fig:dists}d has a clear plateau at velocities $-5500<v_{||}<-3000$ km/s. Further away from the X-line, Fig.~\ref{fig:dists}e, the beam becomes faster and narrower in energy. Finally, a close to symmetric distribution (Fig.~\ref{fig:dists}f) is observed, which is characteristic for the magnetospheric inflow region \citep{egedal2011,graham2016}.

Compared to the magnetosheath inflow, the electrons within the jet are significantly hotter, Fig.~\ref{fig:waves}c. $T_{e||}$ increases by a factor of 2.5 (max), and  $T_{e\perp}$ by a factor of 2. The  $T_{e\perp}$ increase cannot be attributed to adiabatic betatron heating alone ($T_{\perp a}$ in Fig.~\ref{fig:waves}c), indicating non-adiabatic heating. We note that the peak of the beam in Fig.~\ref{fig:waves}b follows closely $-v_{Te||}$, which indicates that the energy gained through the accelerating potential $\Delta \Phi_{||}$ (discussed later) is transformed into the parallel heating gradually. 

To investigate the possible physical mechanism responsible for the heating, we look into waves as
the jet region is abundant with E and B fluctuations. Magnetic fluctuations are confined to $f<f_{ce}$ (Fig.~\ref{fig:waves}d) and have primarily right-hand polarisation close to circular (ellipticity $\sim$1, Fig.~\ref{fig:waves}e), indicating whistler-mode waves. 
For a spectral peak at 200 Hz$\lesssim0.25 f_{ce}$, we find wave-normal angles $\theta_k < 20^\circ$, which correspond to a quasi-parallel whistler.  For $f>0.25 f_{ce}$ the B power drops significantly (Fig.~\ref{fig:waves}d), and the fluctuations become more electrostatic with more oblique wave vectors.
Electric field fluctuations reach up to $f \sim f_{pe}$, and have an intermittent character and generally broadband spectrum. $E_{||}$ fluctuations dominate for $f>f_{ce}$, and $E_{\perp}$  fluctuations are significant for $f<f_{ce}$. Large amplitude $E_{||}$ bursts coincide with the region where plateaus in $f_e(v_{||})$  are observed (Fig.~\ref{fig:waves}b), suggesting a connection between the two.

Figure \ref{fig:dists}h shows $E$ waveforms in the EDR vicinity. We separate the waveforms into the low- and high-frequency (LF and HF) components by low/high-pass filtering at 2.7 kHz. The LF waves in the first half of the interval have polarization close to linear. As the waves are electrostatic, the maximum variance of $\mathbf{E}$ gives $\theta_k = 50^\circ$. For these waves $E_{||} \simeq E_{\perp 2} \gg E_{\perp 1}$, where the  $\perp 1$ direction is approximately normal to the boundary, thus $\mathbf{k}$ is located in the ML plane.  A hodogram corresponding to the wave burst is shown in  Fig.~\ref{fig:dists}i. Given the magnetic field polarization discussed above, we interpret these waves as oblique quasi-electrostatic whistlers. So, both quasi-parallel and obliques whistlers are observed in the same region. WHAMP \citep{ronnmark83} analysis based on distribution Fig.~\ref{fig:dists}e shows that the oblique whistler is generated by the electron beam via Landau resonance, while the quasi-parallel whistler is generated by perpendicular temperature anisotropy created due to spreading of the beam in  $v_{\perp}$ as it propagates towards the stronger magnetic field region. In this case, the generation of parallel whistlers is different from the other magnetopause/separatrix  cases, where whistlers are generated by the loss-cone distributions produced by the escape of magnetospheric electrons along newly-opened field lines \citep{graham2016}.  
The quasi-parallel whistlers can possibly contribute to the observed non-adiabatic increase of $T_{e\perp}$ (Fig.~\ref{fig:waves}c). 

In the second part of the interval, in the EDR vicinity, the LF waves  as well as the HF waves have $E_{||} \gg E_{\perp}$.
Here, the magnetic field is close to the spacecraft spin plane, which allows usage of the 120-meter-long SDP booms for inter-probe interferometry \citep{graham2015} to estimate velocity of the $E_{||}$ structures, $v_{ph}$. The waves are electrostatic, and have $\mathbf{k}$ aligned with $\mathbf{B}$ based on the maximum variance analysis of $\mathbf{E}$, Fig.~\ref{fig:dists}j,k. We find that the LF and HF waves have distinct speeds. LF waves (e.g., Fig~\ref{fig:dists}k) propagate anti-field-aligned (in the electron flow direction) with speeds in the range 150-300 km/s in the ion frame. Errors in the $v_{ph}$ estimates are below 30\% \citep{steinvall2019}. The HF waves are $\sim$10 times faster. The obtained  $v_{ph}$ correspond to wavelength of $\sim$10-20 $\lambda_D$ for both LF and HF waves, where $\lambda_D$ is the Debye length. We interpret the slow LF waves as Buneman mode and the fast HF waves as the beam mode. This is supported by WHAMP analysis using a model distribution based on the observation, which consists of the electron jet and the hot magnetospheric background. The existence of the background enables both the beam modes and the Buneman mode generation.

\begin{figure}
\includegraphics[width=8.6cm]{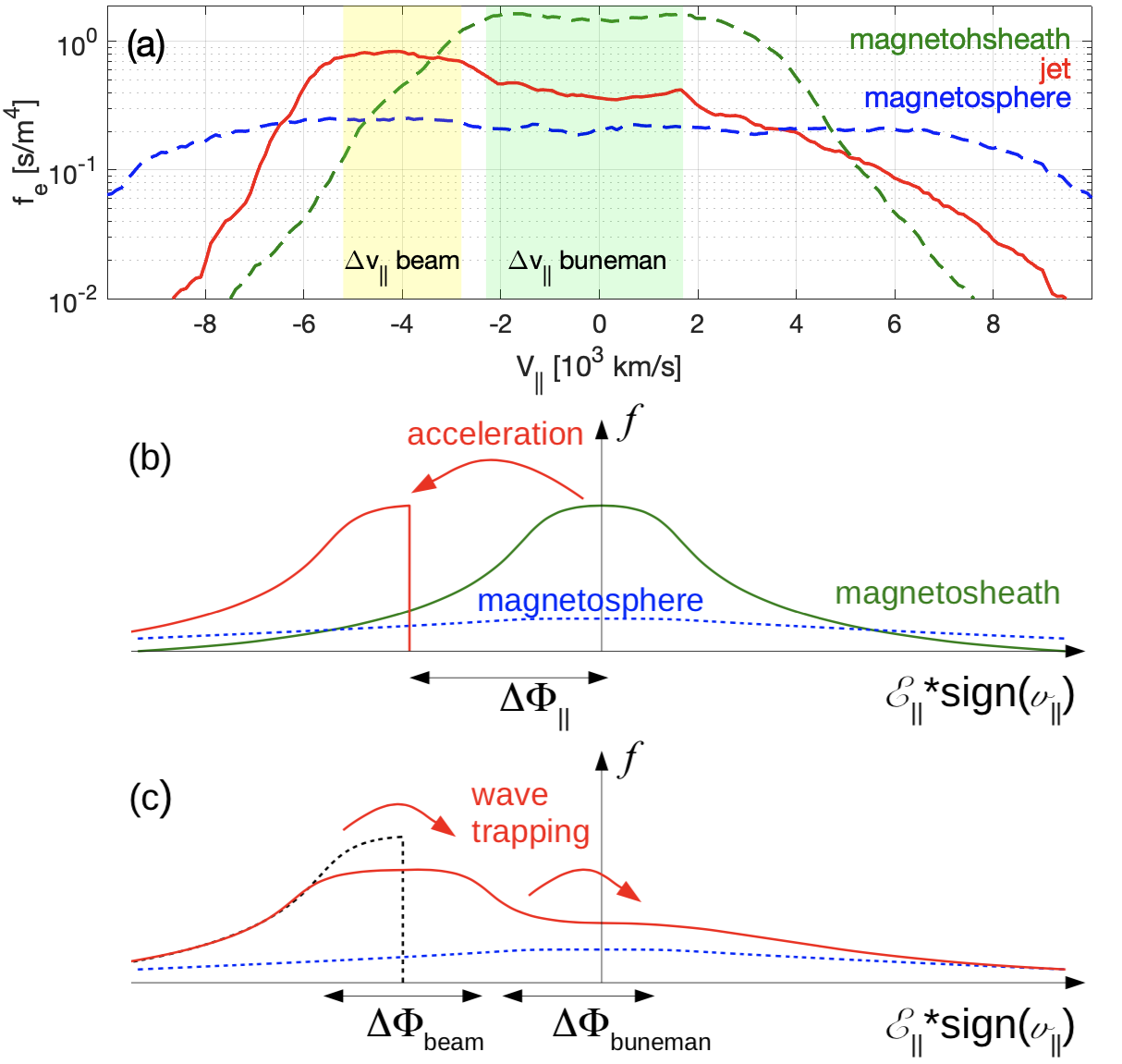}
\caption{Observed VDF of the jet and schematic of electron distribution evolution. (a) integrated 1D VDFs of the electron jet (same as distribution d in Fig~\ref{fig:dists}g) and trapping ranges for the beam and Buneman waves. Magnetosheath and magnetosphere VDFs are provided as reference. The observed jet VDF is interpreted as a result of: (b) initial acceleration by $E_{||}$ (reconnection electric field) followed by
(c) thermalization of the electrons due to interaction with the beam and Buneman modes.
\label{fig:schematic}}
\end{figure}

Using the observed wave amplitudes and the obtained $v_{ph}$ we can evaluate the wave potential $\varphi = - \int E_{||} dl_{||} = \int E_{||} v_{ph} dt$. The interval of $\Delta v_{||}$ in which the finite amplitude wave will interact with electrons is defined as $v_{ph} \pm (2e\varphi/m_e)^{1/2}$. $\Delta v_{||}$ for Buneman and beam mode waves based on the maximum wave amplitudes are shown in Fig.~\ref{fig:schematic}a. One can see that the two trapping intervals correspond to plateaus in the VDF. Buneman waves have insufficient amplitude to directly trap the electron jet. However, the gap between the two trapping intervals is very small, which suggests that the intervals may at times overlap, and the fastest electrons initially interacting with the beam mode can eventually move to the trapping region of the Buneman wave.    
 
Our interpretation of the observed process is summarized in Fig.~\ref{fig:schematic}. The electron reconnection jet is dominated by the magnetosheath electrons, because of the high density asymmetry. The electrons are accelerated by $E_{||}$ both in the separatrix regions \citep{egedal2009} and at the X-line (reconnection electric field), gaining a substantial potential $\Delta \Phi_{||} \sim80 $ V $\sim T_e$, Fig.~\ref{fig:schematic}b. Acceleration continues until the jet becomes unstable to current-streaming instabilities. Fast beam-driven and slow Buneman waves are generated close to X-line and transform the beam into a plateau, i.e. slow down part of the beam (beam relaxation), Fig.~\ref{fig:schematic}c. Slower Buneman waves have insufficient amplitude to trap the initial fast beam, but they trap the low-energy part of the plateau produced by the fast waves, forming another plateau around zero velocity. The slow velocity of the Buneman waves, $v_{ph}\sim v_{Ti}$, allows coupling of the electron jet to ions and thus can provide anomalous drag \citep{che2017}. Interplay between fast beam-driven and slow Buneman waves is responsible for thermalization of the beam, i.e. initial kinetic energy of the accelerated cold electron jet is transferred into thermal energy. This process results in fast and efficient electron heating via irreversible phase-mixing.


Our study shows that large-amplitude Debye-scale electrostatic turbulence is generated at the X-line of asymmetric reconnection with moderate guide field, and this turbulence has a strong effect on the electron jet evolution supporting earlier theoretical predictions \citep{chen2009,munoz2018}. While for antiparallel  and weak guide-field reconnection the electron dynamics in the EDR is largely laminar and dominated by the meandering electron orbits, for stronger guide field configurations, which are common in astrophysical plasma environments, electrostatic turbulence may play a major role. 

\begin{acknowledgments}
We thank the entire MMS team for data access and support. This work was supported by the Swedish National Space Agency, grant 128/17, and the Swedish Research Council, grant 2016-05507. MMS data was accessed from https://lasp.colorado.edu/mms/sdc/public on 20 November 2019. Data analysis was performed using the IRFU-Matlab analysis package available at https://github.com/irfu/irfu-matlab.
\end{acknowledgments}

\end{document}